\documentclass[twocolumn, secnumarabic, amssymb, amsmath,nobibnotes, aps, prl,nofootinbib]{revtex4}

\usepackage{graphicx}

\date{\today}

\newcommand{\half}{\mbox{\small{$\frac{1}{2}$}}}

\newcommand{\baru}{\bar{u}}
\newcommand{\barv}{\bar{v}}
\newcommand{\barw}{\bar{w}}

\begin{document}
\title{aberration and radiation pressure in the Klein and Poincar\'e models}
\author{B. H. Lavenda}
\email{bernard.lavenda@unicam.it}
\affiliation{Universit$\grave{a}$ degli Studi, Camerino 62032 (MC) Italy}
\begin{abstract}
Aberration and radiation pressure reflected by a moving mirror are examples of the Klein, one-way Doppler shift, and Poincar\'e, two-way Doppler shift, disc models of hyperbolic geometry, respectively. Aberration, like the Thomas precession, is related to the angular defect, and is a kinematical effect rather than relativistic. At the angle of parallelism, determined by a stationary observer looking at a moving object in the direction normal to its motion,  the rotation of the object is related to its Lorentz contraction that an observer sees  traveling at the same speed as the object. The origin of the Lorentz contraction is the angular defect, while the angle of parallelism is an asymptotic limit, providing the unique link between circular and hyperbolic functions. The relative velocity provides an upper limit on the angle of incidence with the radiation pressure vanishing at the  angle of parallelism.  Two-way, second-order Doppler shifts can be used to establish experimentally the existence of an angle of parallelism. \end{abstract}

\maketitle

\begin{quote}
The hyperbolic distance ($h$-distance) in the Klein model \lq\lq differs from the formula in the Poincar\'e disc model by a mere factor of two!\rq\rq~\cite{needham}
\end{quote}

\section{Angular defect and its relation to aberration and Thomas precession}
Although the setting of hyperbolic geometry for relativity in general~\cite{fock}, and special relativity in particular~\cite{varicak1,arnold,prokhovnik}, is not new, what is new are the physical predictions which can be drawn from it. Fock~\cite{fock} showed that Friedmann's~\cite{friedmann} solution to Einstein's equation, corresponding to a uniform mass density at zero pressure, can be formulated in Lobachevsky's velocity space, which has become known as \lq rapidity\rq\ space~\cite{sard}. 

The angular defect concerns both aberration and parallax, although the two phenomena are quite distinct from one another~\cite{sommerfeld}. In fact, Bradley discovered aberration in 1728 while looking for parallax. Although both phenomena cause the locus of a star to trace out an ellipse, the direction and magnitude of the angular deviation is quite different from that caused by parallax. The crucial difference is that the magnitude of deviation caused by aberration is independent of the distance to the star, and is much greater than for parallax. It is known that the angle of parallax is greater than the defect~\cite{faber}, and, moreover, the angle of parallax is greater than the complementary angle of parallelism, which is a sole function of distance. In the Klein model, we will appreciate that the angle of parallelism is a limiting angle, while the angular defect is always present.

It has also been shown that the angular defect of a hyperbolic triangle is related to the upper bound on the Euclidean ($e$-) measure of relativistic velocities using the Poincar\'e disc model which is conformal~\cite{criado}. On the other hand, if the Klein model is used, which is not conformal, one would find Lorentz contraction in the direction normal to the motion~\cite{lavenda}.  

The angular defect in the hyperbolic triangle, which is proportional to the area, has also been implicated in the  determination of the rotation of axes in successive Lorentz transformations  in different planes~\cite[pp.273-281]{sard}. It came as a curious surprise that successive Lorentz transforms, or \lq boosts\rq\ as they are now referred to, is not another boost, but involves a rotation. In physics, the angle of rotation is known as Wigner's angle~\cite{aravind}, and is the kinematic factor underlying Thomas precession.

 If $\vec{u}$ and $\vec{v}$ are two velocities then the most general composition law is~\cite{busemann}:
\begin{equation}
w=\frac{\surd[(\vec{u}-\vec{v})^2-(\vec{u}\times\vec{v})^2/c^2]}{1-\vec{u}\cdot\vec{v}/c^2}. \label{eq:comp}
\end{equation}
 The non-planar aspects of the composition law can be clearly seen in the second term of the numerator of \eqref{eq:comp}. Expression \eqref{eq:comp} can also be derived by differentiating the Lorentz transformations at constant, relative velocity~\cite[pp. 46-47]{fock}. Then, introducing $\vec{v}=\vec{u}+d\vec{u}$ into \eqref{eq:comp}, and dividing through by $dt$,  the law of acceleration is obtained as:
 \begin{equation}
 \dot{w}=\frac{\surd[\dot{u}^2-(\vec{u}\times\dot{\vec{u}})^2/c^2]}{1-u^2/c^2}. \label{eq:a}
 \end{equation}
 
 This decomposes the acceleration into longitudinal ($\vec{u}\parallel\dot{\vec{u}}$), and transverse components, analogous to longitudinal and transverse masses. It is the second term in the numerator of \eqref{eq:a} that is related to the Thomas precession: the rotation of the electron's velocity vector~\cite[p. 286]{sard}, 
 \begin{equation}
 d\vartheta=(\vec{u}\times\dot{\vec{u}})dt/u^2,\label{eq:rot}
 \end{equation} caused by the acceleration, $\dot{\vec{u}}$, in time, $dt$. Then, as the velocity turns by $d\vartheta$ along the orbit, the spin projection turns in the opposite direction by an amount equal to the angular defect of the hyperbolic triangle whose vertices are the velocities in three different inertial frames in pure translation with respect to one another. 
 
 The defect caused by aberration can be readily calculated. Consider the triangle formed by three vertices $\vec{u}_1$, $\vec{u}_2$, and $\vec{u}_3$ in velocity space. By setting $\vec{u}_3=\vec{n}c$, where $\vec{n}$ is the unit normal in the direction of the light source, we are considering an ideal, or \lq improper,\rq\ triangle~\cite{kul}, which shares many properties of ordinary triangles, but has the property that the sum of its angles is less than two-right angles, its so-called defect. Consequently, there will be two parallel lines forming an ideal vertex $\vec{u}_3$, whose angle is zero so that $\cos\vartheta_3=1$.  
 
The cosines of the angles are given by the inner products~\cite{busemann}:
 \begin{equation}
\cos\vartheta_i= \frac{(\vec{u}_k-\vec{u}_i)\cdot(\vec{u}_j-\vec{u}_i)-(\vec{u}_k\times\vec{u}_i)\cdot(\vec{u}_j\times\vec{u}_i)/c^2}{\Delta_{ik}\;\Delta_{ij}}, \label{eq:cos}
\end{equation}
where $\Delta_{ik}=\surd[(\vec{u}_k-\vec{u}_i)^2-(\vec{u}_k\times\vec{u}_i)^2/c^2]$, and a similar expression for $\Delta_{ij}$. All three angles can be calculated by permuting cyclically the indices, and it is easy to see that $\vartheta_3=0$.
By choosing a frame where the velocities are equal and opposite in direction, $\vec{u}_1=-\vec{u}_2$, we are, in fact, considering a \lq two-way\rq\ Doppler shift. The relative velocity is:
\begin{equation}
\gamma=\frac{2\beta}{1+\beta^2}, \label{eq:gamma}
\end{equation}
where $\beta=u/c$, and $u=|\vec{u}_1|=|\vec{u}_2|$. The projection of the velocity on the normal to the wavefront is: 
\begin{equation}
\vec{n}\cdot\vec{u}_1=-\vec{n}\cdot\vec{u}_2=u\cos\vartheta\ge0. \label{eq:ineq}
\end{equation}

The cosine law \eqref{eq:cos} for angles $\vartheta_1$ and $\vartheta_2$ can be written as:
\begin{equation}
 \cos\vartheta_i=\frac{(u^2-c\vec{n}\cdot\vec{u}_i)}{u(c-\vec{n}\cdot\vec{u}_i)}\;\;\;\; i=1,2. \label{eq:Cos}
 \end{equation}
On account of \eqref{eq:ineq}, the cosine of the first angle is:
 \begin{equation}
 \cos\vartheta_1=\frac{\beta-\cos\vartheta}{1-\beta\cos\vartheta}. \label{eq:cos1}
 \end{equation}
 This represents the usual formula for aberration, except for the negative sign which implies reflection and guaranteeing that the angle of parallelism is acute. The second equation of aberration for the first angle is:
 \begin{equation}
 \frac{\sin\vartheta_1}{\lambda_1}=\frac{\sin\vartheta}{\lambda}. \label{eq:mirror}
 \end{equation}
 The ratio of the wavelengths,
 \begin{equation}
 \frac{\lambda_1}{\lambda}=\frac{\surd(1-\beta^2)}{1-\beta\cos\vartheta}, \label{eq:christian}
 \end{equation}
 is Doppler's principle. For the second angle we have:
 \begin{equation}
 \cos\vartheta_2=\frac{\beta+\cos\vartheta}{1+\beta\cos\vartheta}, \label{eq:ab1}
 \end{equation}
 again on account of \eqref{eq:ineq}, and, hence,
 \begin{equation}
  \sin\vartheta_2=\frac{\surd(1-\beta^2)}{1+\beta\cos\vartheta}\sin\vartheta.\label{eq:ab2}
  \end{equation}
  Finally, by \eqref{eq:ineq}, we find the relation:
  \begin{equation}
  \cos\vartheta_2=\frac{\gamma-\cos\vartheta_1}{1-\gamma\cos\vartheta_1}, \label{eq:Cos3}
  \end{equation}
 between the two cosines, where $\gamma$ is the relative speed given by \eqref{eq:gamma}.
 
The aberration formula \eqref{eq:mirror} has the identical form of the law of reflection for a moving mirror~\cite{kennard}. For a stationary mirror, $\lambda_1=\lambda$ and $\vartheta_1=\vartheta$, where the angles are subtended by the incoming and outgoing rays, and the surface of the mirror. However, it must be borne in mind that the angles are at the vertices in velocity space so that an angle of $\vartheta=\pi/2$ is parallel to the wavefront, or perpendicular to the motion.
 
  The first, \eqref{eq:cos1} and \eqref{eq:mirror}, and second, \eqref{eq:ab1} and \eqref{eq:ab2}, pair of aberration equations can be combined to read:
\begin{subequations}
\begin{align}
\tan(\vartheta_1/2)=\left(\frac{1-\beta}{1+\beta}\right)^{1/2}\cot(\vartheta/2), \label{eq:ab-cot}\\
\tan(\vartheta_2/2)=\left(\frac{1-\beta}{1+\beta}\right)^{1/2}\tan(\vartheta/2), \label{eq:ab-tan}
\end{align}
\end{subequations}
respectively.
Expression \eqref{eq:ab-tan} is the usual formula given for aberration~\cite{rindlerbis}. By letting the third vertex be the the speed of light, we have formed an ideal triangle. In hyperbolic geometry, a transversal which cuts the two parallel lines forms angles  in the direction of parallelism such that the sum of the angles  is less than two right angles.  

Two important cases arise: (i) when the vertices, $\vec{u}_1$ and $\vec{u}_2$, of the ideal triangle are on the same limiting curve, called a horocycle, $H$, whose center is at infinity,  $\Omega$, shown in Fig. 1, and (ii) when the transversal is perpendicular to one of the parallel lines, shown in Fig. 2. 

\begin{figure*}[floatfix]
	\centering
		\includegraphics[width=0.50\textwidth]{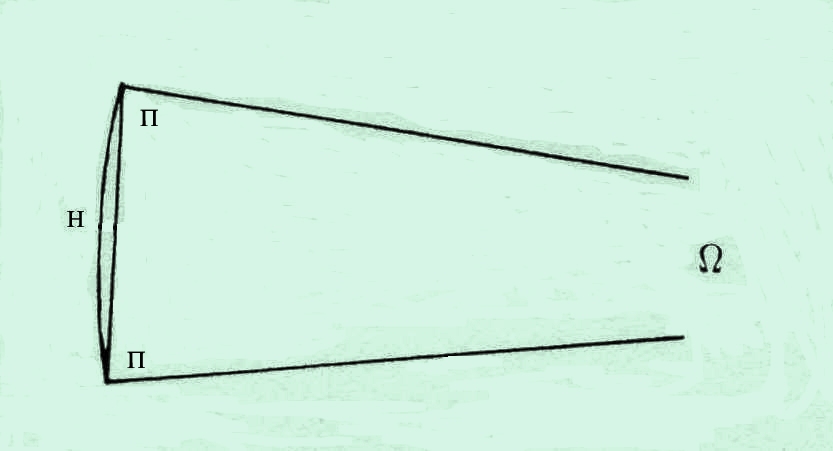}
		\caption{}
	\label{fig:P-1}
\end{figure*}

\begin{figure*}[floatfix]
	\centering
		\includegraphics[width=0.50\textwidth]{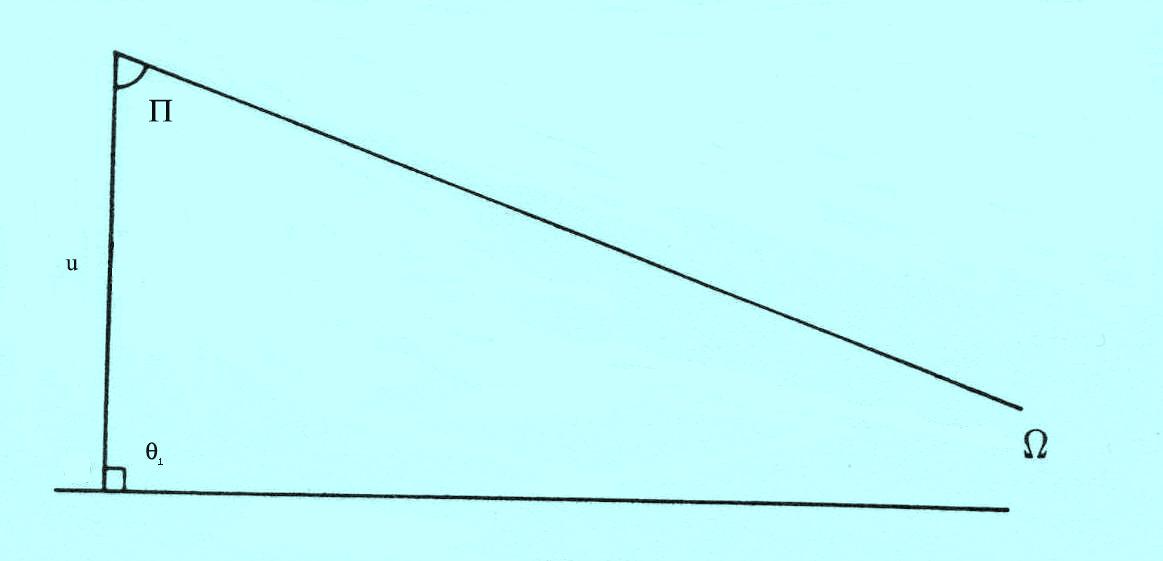}
		\caption{}
	\label{fig:P-2}
\end{figure*}
In the first case, $\vartheta=\pi/2$, $\vartheta_1=\vartheta_2=\Pi$, where $\Pi$, the angle of parallelism, is given by: 
\begin{equation}
\tan(\Pi(u/2)/2)=\left(\frac{1-\beta}{1+\beta}\right)^{1/2}=e^{-u/c}. \label{eq:Tan1}
\end{equation}
$\Pi$  is only a function of the \emph{half\/} \lq distance\rq\ $u/2$.  From \eqref{eq:cos1} and \eqref{eq:ab1} we find $\cos\vartheta_1=\cos\vartheta_2=\beta$, which is the $h$-measure of distance in velocity space. 

In the second case, one of the angles is $\pi/2$, and the other is necessarily acute, being the angle of parallelism. In other words, lines with a common perpendicular cannot be parallel so that $\Pi$ must be acute. It is readily  seen from \eqref{eq:ab-tan} that $\vartheta_2$ cannot become a right-angle because that would imply $\cos^{-1}(-\beta)=\vartheta>\pi/2$, and so violate \eqref{eq:ineq}. $\beta$ is the $e$-measure of length, which is equal to the hyperbolic tangent of its $h$-measure [cf. eqn \eqref{eq:tanh} below]. Negative values are ruled out in hyperbolic geometry: \lq\lq the hyperbolic tangent is a function that assumes all values between $0$ and $1$\rq\rq~\cite[p. 163]{kul}. In other words, the angle of parallelism must be an acute angle, for, otherwise, the lines would be divergent. The formation of an ideal triangle is related to the fact that $c$ is the limiting speed. We will return to this point when discussing Terrell's \lq\lq Invisibility of the Lorentz contraction.\rq\rq\

Rather, if $\vartheta_1=\pi/2$, and  \eqref{eq:ab-cot} is introduced into \eqref{eq:ab-tan}, we get:
\begin{equation}
\tan(\Pi(u)/2)=\frac{1-\beta}{1+\beta}=e^{-2u/c}, \label{eq:Tan2}
\end{equation}
where the angle of parallelism, $\Pi$, is a function of the \emph{whole\/} \lq distance\rq\ $u$.
From \eqref{eq:Cos3} we find the new $h$-measure of distance as $\cos\vartheta_2=\gamma$, which again related to the hyperbolic tangent through  \eqref{eq:iso-bis} below. In contrast to \eqref{eq:Tan1}, the $h$-measure has become twice as great in \eqref{eq:Tan2}. This, as we shall see, is the same as performing a \lq two-way\rq\ Doppler shift.

The defect, $\eta=\pi-\theta_1-\theta_2>0$, is expressed in terms of the relative speed $\beta$ and the angle $\vartheta$ subtended by the direction of the light source and the line of sight of the observer, i.e.,~\cite[p. 53]{fock}:
\begin{equation}
\tan(\eta/2)=\frac{\beta}{\surd(1-\beta^2)}\sin\vartheta. \label{eq:defect}
\end{equation}

In Thomas procession, the velocity turns along the orbit by an amount $\vartheta$, while the spin projection in the orbital plane turns in the opposite direction by the amount $\eta$, the $h$-defect~\cite[p. 288]{sard}. Under the same conditions as above, the angular defect has been found to be related to the rotation of the velocity vector \eqref{eq:rot} by:
\begin{equation}
\sin(\eta/2)=\half(\Gamma-1)\sin(d\vartheta), \label{eq:thomas}
\end{equation}
where $\Gamma$ is not the Lorentz factor, as usually assumed~\cite[p. 289]{sard}, but, rather is~\cite{rhodes}:
\begin{equation}
\Gamma=\left(1-\gamma^2\right)^{-1/2}=\frac{1+\beta^2}{1-\beta^2}. \label{eq:Gamma}
\end{equation}
 For small defects where the circular functions on the left-hand sides of \eqref{eq:defect} and \eqref{eq:thomas} can be approximated by their arguments, the proportionality factor to the sine of the angle of rotation of the velocity vector in \eqref{eq:thomas} is the \emph{square\/} of $\beta/\surd(1-\beta^2)$ in \eqref{eq:defect}. This tends to imply that a \lq two-step\rq\ process is involved where the Doppler shifts in \eqref{eq:ab-tan} and \eqref{eq:ab-cot} are squared [cf. eqn \eqref{eq:tan-bis} below]. In that case, we obtain
 \begin{equation}
 \tan(\eta/2)=(\Gamma-1)\sin\vartheta, \label{eq:defect-bis}
 \end{equation}
 instead of \eqref{eq:defect}.
For small defects,  \eqref{eq:defect-bis}  is seen to be a factor of $2$ greater than \eqref{eq:thomas}, if the Lorentz factor is given by compounded expression \eqref{eq:Gamma}.~\footnote{In fact, expression \eqref{eq:defect-bis}  is precisely the result found in Ref.~\cite{rhodes}, eqn (119), when eqn (115) is introduced, and their $d\chi$ is set equal to $\eta/2$, where the defect, $\eta$, is the negative of the rotation of the spin projection in time $dt$. In their eqn (116), they identify $\vec{r}$ with $\vec{v}$. Their eqn (67) is an identity by identifying $r$ with $v$, $\gamma$ with $\Gamma$, and their $\beta$ with $\gamma$. However, their eqn (120) does not follow from eqn (119).} 

Moreover, either expression \eqref{eq:defect}, or \eqref{eq:defect-bis}, shows that the turning of the spin is kinematical in origin, and not relativistic as it would be if $\Gamma$ were the \lq one-way\rq\ Lorentz factor, $(1-\beta^2)^{-1/2}$, since \lq\lq $(1-\beta^2)^{-1/2}-1\approx0$  in the low velocity region\rq\rq~\cite[p. 289]{sard}. Any time a component of the acceleration exists normal to the velocity, \lq\lq for whatever reason, then there is a Thomas precession, independent of other effects\rq\rq~\cite{jackson}---including relativistic ones. This kinematical effect is due to the compounding of Doppler shifts, and will be a recurrent theme throughout this paper.

The fundamental connection between hyperbolic geometry and optical phenomena in general, and relativity in particular, is that the longitudinal Doppler shift is related to the exponential distance of the Klein model in velocity space~\cite{lavenda}.  Hyperbolic length is defined through the cross-ratio; the cross-ratio is a projective invariant of four points. This is the smallest number of points that is invariant, since three points on a line may be projected to any other three. Whereas the Klein disc is not conformal,  except at the origin of the hyperbolic plane,  the Poincar\'e disc is. The disc models also differ in how $h$-distance is measured:  the $h$-distance is twice as great in the Poincar\'e disc than it is in the Klein disc. The factor two is not just a mere numerical factor, since it is indicative of reflections and the way velocities are compounded and distances measured. Moreover, it will change the dependencies of energy, momentum, and consequently, mass, on the relative speed.

Another possibility of vindicating hyperbolic geometry consists in the distinction between aberration and the pressure of radiation against a moving mirror. Early in the development of special relativity, the Lorentz transform and its inverse were used to determine the pressure of radiation on a moving mirror~\cite{max,einstein}. It is still common to use aberration to determine the radiation pressure, even though Einstein calculated the difference in the energy density \emph{after\/} being reflected from the mirror and the initial energy density in order to determine the radiation pressure. A \lq two-way\rq\ Doppler shift is involved, and not a one-way Doppler shift~\cite{terrell}. This we will show to be the same distinction between the Klein and the Poincar\'e models of hyperbolic geometry. Moreover, it will turn out that the second-order Doppler effect predicted by the two-way Doppler shift is an experimental test for the angle of parallelism.

\section{Isomorphism of the Klein model onto the Poincar\'e model}

The relativistic velocity addition law for two systems moving at equal and opposite speeds, \eqref{eq:gamma}, is the isomorphism from the Klein model of hyperbolic geometry onto the Poincar\'e  disc and upper half-plane models~\cite{greenberg}.  In the Poincar\'e disc model, points of the hyperbolic plane are represented by points interior to an $e$-circle, $\Gamma$.  Lines, not passing through the center of the circle,  are represented by open arcs of circles which cut a fixed circle, $\Gamma$, orthogonally at $P$ and $Q$ in Fig. 3. The only straight lines pass through the center, which are also orthogonal to the unit circle, called the circle at infinity, or the \lq horizon.\rq\
\begin{figure*}[floatfix]
	\centering
		\includegraphics[width=0.50\textwidth]{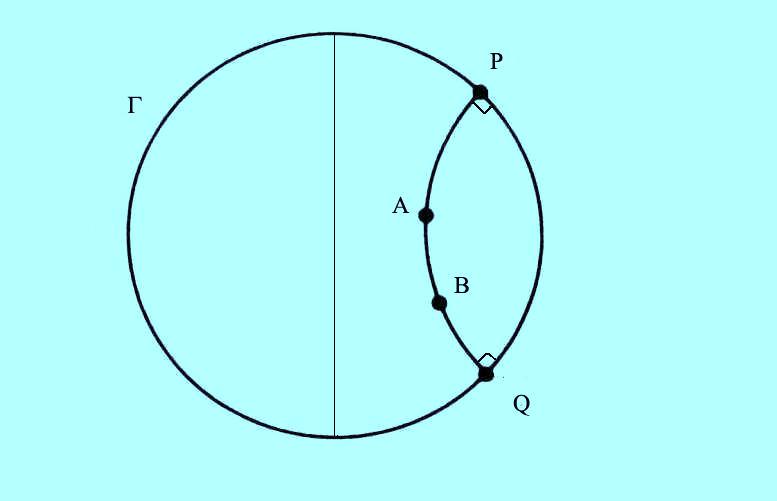}
		\caption{}
	\label{fig:P-3}
\end{figure*}

Not only did Beltrami discover the Poincar\'e disc model, some fourteen years before Poincar\'e rediscovered it~\cite[p. 315]{needham}, he also constructed the Klein, or projective model, by projecting a hemisphere vertically downwards onto the complex plane. Although the projection of a small circle on the hemisphere becomes an ellipse on the disc, so that the Klein model is not conformal, the redeeming virtue of the model is that the vertical sections of the hemisphere are projected into $e$-straight lines as shown in Fig. 4. In other words, the $h$-lines of the Klein model are $e$-chords of the unit circle.
\begin{figure*}[floatfix]
	\centering
		\includegraphics[width=0.50\textwidth]{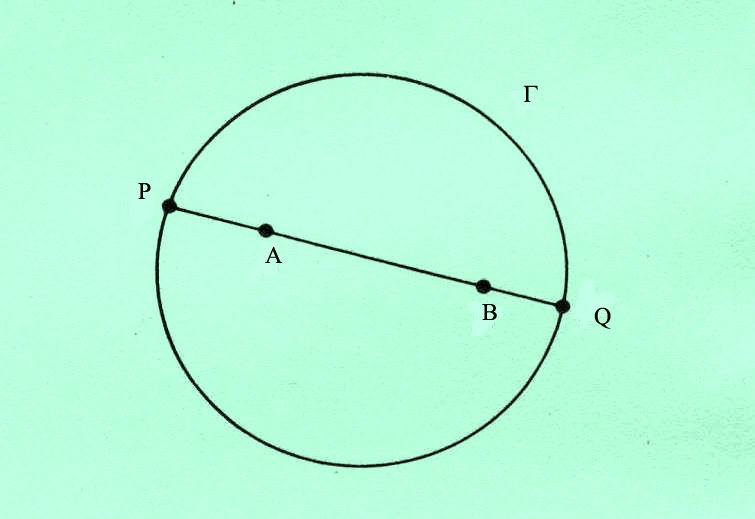}
	\caption{}
	\label{fig:P-4}
\end{figure*}

 If  $A$ and $B$ are \emph{ordinary\/} points inside $\Gamma$, and $P$ and $Q$ are the ends of the  chord through $A$ and $B$, Klein's definition of the length of the segment is:
\begin{equation}
d(AB)=\half |\ln(AB,PQ)|, \label{eq:d-klein}
\end{equation}
where 
\[(AB,PQ)=\frac{\overline{AP}}{\overline{AQ}}\cdot\frac{\overline{BQ}}{\overline{BP}},\]
is the invariant cross-ratio. The overbar represents $e$-length, and we will use it to distinguish from the $h$-length when confusion can arise. Poincar\'e, on the other hand, determines the distance of the arc from $A$ to $B$ as twice Klein's distance, viz.,
\begin{equation}
d^{\prime}(AB)=|\ln(AB,PQ)|. \label{eq:d-poincare}
\end{equation}

Let the ends of the chords $P$ and $Q$ be $1$ and $-1$. If $X$ and $Y$ have coordinates $x$ and $y$ then the cross-ratio is:
\[(XY,PQ)=\frac{1+x}{1-x}\cdot\frac{1-y}{1+y}.\]
If $A=\gamma(x)$ and $B=\gamma(y)$, it follows  that $d(AB)=d^{\prime}(XY)$ since
\[\frac{1+\gamma(x)}{1-\gamma(x)}=\left(\frac{1+x}{1-x}\right)^{2}.\]
Hence, $\gamma$ given by \eqref{eq:gamma}, is an isomorphism that makes the lengths of the Klein and Poincar\'e models to coincide.

\section{aberration versus radiation pressure on a moving mirror}
\subsection{aberration and the angle of parallelism}
Having derived the formulas for aberration in the first section, we now consider, in greater detail, the limiting forms \eqref{eq:Tan1} and \eqref{eq:Tan2} which are the Bolyai-Lobachevsky formulas for the angle of parallelism.
Although there has been no mention of hyperbolic geometry, this situation has been widely discussed in the literature,~\cite{terrell-bis,weisskopf}, and without the connection to an angle of parallelism. 

For $\vartheta^{\prime}=\pi/2$, the observer in a frame in which the object is at rest will see the object rotated by an amount $\sin\vartheta=\surd(1-\beta^2)$, just equal to the Lorentz contraction. The angle of parallelism, $\vartheta=\cos^{-1}\beta$ provides the link between circular and hyperbolic functions. Only at the angle of parallelism can a rotation be equated with a Lorentz contraction. 

Terrell~\cite{terrell-bis} also considers the opposite case where $\vartheta=\pi/2$ and $\vartheta^{\prime}=\cos^{-1}(-\beta)$. He concludes that to the stationary observer, the object appears \lq\lq to be rotating about its line of motion in such a way as to appear broadside at $\vartheta^{\prime}=\cos^{-1}(-\beta)$, and to present a view of its rear end from that time on.\rq\rq\ However, the stationary observer will not see any motion of this sort performed by the moving object because the angle of parallelism,  linking circular and (positive) hyperbolic functions, must be acute;  otherwise, the $h$-measure of distance would turn out to be negative!  Therefore, Terrell's~\cite{terrell-bis,weisskopf}  analysis cannot be extended to angles of parallelism greater than $\pi/2$, for such angles do not exist. In other words, the observer must make his observation of the object in the same inertial frame of the object, and the condition $\vartheta^{\prime}=0$ makes $\vartheta$ an angle of parallelism via the equation of aberration, \eqref{eq:mirror}, or \eqref{eq:ab2}. This is to be contrasted by a comment made by Vari\'cak~\cite{varicak-bis} which provoked Einstein's response~\cite{albert}:~\footnote{It is rather ironic that Vari\'cak's work on hyperbolic geometry went almost completely unnoticed, yet his small note on whether the Lorentz contraction was real or not caused a great deal of commotion and confusion~\cite{miller}.}
\begin{quote}
The question of whether the Lorentz contraction is real or not is misleading. It is not \lq real\rq\ insofar as it does not exist for an observer moving with the object.
\end{quote}

We will analyze the angle of parallelism further in terms of the projective disc model, showing that leads to Lorentz contraction in a direction normal to the motion [cf. eqn \eqref{eq:alpha} below]. Here, we will relate it with the vanishing of the radiation pressure on a moving mirror.

The angle of parallelism in \eqref{eq:Tan1} is a sole function of half the \lq distance\rq\ $u$. The latter is the hyperbolic measure of distance in velocity space,
\begin{equation}
u=\frac{c}{2}\ln\left(\frac{1+\beta}{1}\cdot\frac{1}{1-\beta}\right)=c\tanh^{-1}\beta, \label{eq:u}
\end{equation}
whose $e$-measure is $\baru=\beta c$, and the speed of light, $c$, is the absolute constant of the hyperbolic geometry. 

More precisely, \eqref{eq:u}, is the Klein length of the velocity segment. On the basis of \eqref{eq:u}, we get the basic relation for the measure of a straight line segment in Lobachevsky space
\begin{equation}
\beta=\tanh(u/c)=\cos\vartheta(\baru), \label{eq:tanh} 
\end{equation}
and
\begin{equation}
\cosh(u/c)=\frac{1}{\surd(1-\beta^{2})}\; ;\;\;\;\; \sinh(u/c)=\frac{\beta}{\surd(1-\beta^{2})}. \label{eq:hyper}
\end{equation}
Whereas the first equality in \eqref{eq:tanh} and \eqref{eq:hyper} hold for all one-way Doppler shifts, the second equality in \eqref{eq:tanh} is valid only at the angle of parallelism, where $\vartheta(u)$ is a function only of $u$.

\subsection{reflection from a moving mirror}
If $\vartheta$ is the angle that a ray makes with the surface of a mirror, and $\vartheta^{\prime\prime} $ the angle of the reflected ray with respect to the surface of the mirror then the law of reflection states that $\vartheta=\vartheta^{\prime\prime}$. This changes when the mirror is in motion. The radiation pressure has a long history since Clerk Maxwell first predicted it~\cite{poynting,larmor}. It also constituted one of the early testing grounds of relativity~\cite{max,einstein}.

If the mirror is receding from the radiating source, the ratio of the wavelengths of impinging and reflected radiation is:
\begin{equation}
\frac{\lambda}{\lambda^{\prime\prime}}=\frac{\cos\vartheta+\beta}{\cos\vartheta^{\prime\prime}-\beta} ,\label{eq:lambda}
\end{equation}
because the wavelength is lengthened in the forward direction and shortened in the backward direction. The angle of reflection is referred to the frame in which the source is at rest, 
\begin{equation}
\cos\vartheta^{\prime\prime}=\frac{\cos\vartheta+\gamma}{1+\gamma\cos\vartheta}, \label{eq:cos-bis}
\end{equation}
where $\gamma$, given by \eqref{eq:gamma}, is the isomorphism from the Klein to the Poincar\'e models, involving a two-step  process for carrying a point $\beta$ in the Poincar\'e disc to the  corresponding point $\gamma$ in the Klein model~\cite[p. 330]{needham}.

Introducing \eqref{eq:cos-bis} into the ratio \eqref{eq:lambda} leads to:
\begin{equation}
\frac{\lambda}{\lambda^{\prime\prime}}=\left(\frac{1+\beta^{2}}{1-\beta^{2}}\right)(1+\gamma\cos\vartheta), \label{eq:lambda-bis}
\end{equation}
which clearly shows that the wavelength of the reflected radiation, $\lambda^{\prime\prime}$, has been shortened with respect to the wavelength of the incoming radiation, $\lambda$. In fact, expression \eqref{eq:lambda-bis} is Doppler's principle, \eqref{eq:christian}, obtained by replacing the relative velocity $\beta$ by $-\gamma$.
Introducing \eqref{eq:lambda-bis} into the aberration equation \eqref{eq:mirror}, which just happens to have the same form as the law of reflection from a moving mirror, results in:
\begin{equation}
\tan(\vartheta^{\prime\prime}/2)=\frac{\sin\vartheta^{\prime\prime}}{1+\cos\vartheta^{\prime\prime}}=
\left(\frac{1-\beta}{1+\beta}\right)\tan(\vartheta/2). \label{eq:tan-bis}
\end{equation}
The ratio of the tangents is the \emph{square\/} of that for aberration, \eqref{eq:ab-tan}!

\section{electromagnetic radiation pressure versus that on a moving mirror}

Clerk Maxwell showed that the pressure exerted on a square centimeter by a beam of light is numerically equal to the energy in a cubic centimeter of the beam. Consider a plane wave of monochromatic light traveling in the $x$-direction. Maxwell's equations for the relevant components of the electric, $E$, and magnetic, $H$, fields are:
\begin{eqnarray*}
E_{x}^{\prime} &=&E_{x}\\
E_{y}^{\prime}&=&\frac{E_{y}-\beta H_{z}}{\surd(1-\beta^{2})}\\
H_{z}^{\prime}&=&\frac{H_{z}+\beta E_{y}}{\surd(1-\beta^{2})},\\
\end{eqnarray*}
where for a plane wave propagating in the $x$ direction, $E_{y}=H_{z}$. The radiation pressure, $P^{\prime}$, in the frame moving at velocity, $\baru$, is related to the pressure in the stationary frame, $P$, according to~\cite{abraham}:
\begin{equation}
P^{\prime}=\frac{1}{2\pi} E_{y}^{\prime\;2}=P\left(\frac{1-\beta}{1+\beta}\right),\label{eq:P}
\end{equation}
where $P=(1/2\pi) E_{y}^{2}$, is Maxwell's prescription of associating the pressure acting on a square centimeter of surface with the energy density in a cubic centimeter of the beam. 

The relativistic Doppler shift in the frequency $\nu^{\prime}$, from its stationary value, $\nu$,
\begin{equation} 
D:=\frac{\nu^{\prime}}{\nu}=\frac{1-\beta\cos\vartheta}{\surd(1-\beta^{2})}, \label{eq:doppler}
\end{equation}
combines the ordinary Doppler shift with the relativistic time dilatation factor. Of course, \eqref{eq:doppler} can be derived from the Lorentz transformation~\cite{becker}; it can also be derived, however, in more general terms from relative velocity, $\barw$,  of the corresponding segment $s$ of the Lobachevsky straight line \eqref{eq:comp}, where the relative velocity is related to the corresponding segment $s$ of the Lobachevsky straight line by $\barw=c\tanh s$. Expression \eqref{eq:comp} spans the entire gamut: from a single velocity, $\beta=c\tanh(u/c)$ [the first equality in eqn \eqref{eq:tanh}], to equal and opposite velocities, $\gamma=c\tanh s$ [eqn \eqref{eq:iso} below]. 

If the energy increases with speed $\barw$ as $E^{\prime}/E_{0}=1/\surd(1-\barw^{2}/c^{2})$,~\footnote{An early contender for the dependency of the energy on the speed was Abraham's~\cite{abe} model which took Searle's~\cite{searle} expression for the total energy of a spherical body of radius $r$ with a uniform distribution of charge, $e$, in motion with a uniform speed $\barw$, \[E=\frac{e^2}{2r}\left\{\frac{c}{\barw}\ln\left(\frac{1+\barw/c}{1-\barw/c}\right)-1\right\},\] 
for the energy of an electron. This expression shows that the energy is  proportional to the difference in the $h$-and $e$-measures of the speed, $(w-\barw)$, where Poincar\'e's $h$-measure is given by the logarithm of the cross-ratio, $w=c\ln[(1+\barw/c)/(1-\barw/c)]$. It  demonstrates that the body's energy, and hence its mass, increases as a result of the motion, and shows that such a dependency depends on the deviation from Euclidean geometry. In the low velocity limit, the energy becomes:
\[E\simeq\frac{e^2}{2r}\left(\frac{1+\bar{w}/c}{1-\bar{w}/c}\right)=\frac{e^2}{2r}e^{w/c},\]
and the momentum is $E/c$.}   then
\begin{equation}
E^{\prime}=E\frac{(1-\baru\cdot\barv/c^{2})}{\surd(1-\baru^{2}/c^{2})}, \label{eq:E}
\end{equation}
where $E/E_{0}=1/\surd(1-\barv^{2}/c^{2})$. For $\barv=c\cos\vartheta$ we get:
\begin{equation}
E^{\prime}=E\frac{(1-\beta\cos\vartheta)}{\surd(1-\beta^{2})}. \label{eq:doppler-E}
\end{equation}
The energy, \eqref{eq:doppler-E}, and amplitude [cf. eqn \eqref{eq:A} below], transform the same way as frequency, \eqref{eq:doppler}. This was stressed by Einstein~\cite{einstein} as being of particular relevance since, according to him, Wien's law is related to it. Since the volume transforms as the inverse of the frequency, the energy density, $\varepsilon$, will transform as the square of the frequency:
\begin{equation}
\varepsilon^{\prime}=D^{2}\varepsilon. \label{eq:epsilon}
\end{equation}

Observing the motion in the line of sight, \eqref{eq:epsilon} reduces to Abraham's expression \eqref{eq:P} for the energy densities. In the general case, the radiation falls obliquely on the mirror, making an angle $\vartheta$ with the normal. A unit area of the mirror will be $1/\cos\vartheta$ times a unit area upon which the rays are falling perpendicularly. In addition, the component of the momentum is reduced by a factor of $\cos\vartheta$ than if it were directed normal to the surface. Consequently, the momentum per unit area is decreased by a factor of $\cos^{2}\vartheta$, and this factor must be multiplied to the energy density when calculating the pressure~\cite{kennard}. Terrell~\cite{terrell} thus obtains:
\begin{equation}
P^{\prime}=2\varepsilon^{\prime}\cos\vartheta^{\prime}=2\varepsilon\frac{\left(\cos\vartheta-\beta\right)^{2}}{\left(1-\beta^{2}\right)}, \label{eq:terrell}
\end{equation}
for the radiation pressure, where the $2$ comes from the fact that, upon reflection, the mirror receives a \lq double dose\rq\ of momentum and the pressure is doubled~\cite[p. 32]{poynting}. 

Whereas the derivation the radiation pressure on a  moving mirror based on aberration is conceptually incoherent~\cite{terrell},  Einstein~\cite{einstein} original derivation is coherent. From his  two-way Doppler shift, and his requirement to calculate the reflected energy in the same frame as the incident energy, he could have deduced many of the results present here, and realized the intimate relationship between relativity and hyperbolic geometry that applies to relativity in general~\cite{lavenda}. We shall show that whereas \eqref{eq:ab-tan} is related to one-way aberration,  its square, \eqref{eq:tan-bis},  relates to the change in wavelength on reflection from a moving mirror.  

Now, Einstein~\cite{einstein} gets the same result as \eqref{eq:terrell}, but uses energy conservation and transforms to the mirror's moving frame, reflects and transforms back to the stationary frame. The first step would have yielded half the pressure, as shown below, but is more enlightening than the method used above, since it brings out the fact that it is  a second-order relativistic effect.  

Einstein obtains the frequency shift after reflection as:~\footnote{Einstein later corrects the denominator to read as in expression \eqref{eq:doppler2}.}
\begin{equation}
\nu^{\prime\prime}=\nu\left(\frac{1+\beta^{2}-2\beta\cos\vartheta}{1-\beta^{2}}\right) ,\label{eq:doppler2}
\end{equation}
which is not the Doppler shift \eqref{eq:doppler}, but, rather, \eqref{eq:lambda}. Moreover, he gives the law of the transform of the cosine of the angle as:
\begin{equation}
\cos\vartheta^{\prime\prime}=\frac{(1+\beta^{2})\cos\vartheta-2\beta}{1+\beta^{2}-2\beta\cos\vartheta}, \label{eq:cos2}
\end{equation}
 which is not the aberration formula \eqref{eq:cos}, but, rather, \eqref{eq:cos-bis} with $\beta\rightarrow-\beta$.  If Einstein used the above procedure to calculate the radiation pressure, he would have obtained:
 \begin{eqnarray}
P& = & 2\varepsilon\left(\frac{1+\beta^{2}-2\beta\cos\vartheta}{1-\beta^{2}}\right)^{2}\left(\frac{(1+\beta^{2})\cos\vartheta-2\beta}{1+\beta^{2}-2\beta\cos\vartheta}\right)^{2}\nonumber\\
&=&2\varepsilon\left(\frac{1+\beta^{2}}{1-\beta^{2}}\right)^2(\cos\vartheta-\gamma)^{2} ,\label{eq:lav}
\end{eqnarray}
which is certainly  not \eqref{eq:terrell}.  This is the radiation pressure that a mirror feels when it moves at constant relative speed $\gamma$.  

Pauli~\cite{pauli} uses the fact that the amplitudes, $A^{\prime}$ and $A$, transform as the frequencies, i.e.,
\begin{equation}
A^{\prime}=A\frac{(1-\beta\cos\vartheta)}{\surd(1-\beta^{2})}, \label{eq:A}
\end{equation}
to claim that the radiation pressure is invariant:
\begin{equation}
P=2A^{2}\frac{(\cos\vartheta-\beta)^{2}}{1-\beta^{2}}=2A^{\prime\;2}\cos^{2}\vartheta^{\prime}=P^{\prime}. \label{eq:P-inv}
\end{equation}
This is not, however, what one would conclude from \eqref{eq:P}.
The invariance of the pressure was first established by Planck~\cite{planck}  by studying how thermodynamic densities transform under the Lorentz transformation. Since $1\ge\cos\vartheta\ge\beta$, we average \eqref{eq:P-inv} over the solid angle with the given limits to get:
\begin{eqnarray}
P_{\mbox{\tiny tot}}(\beta)&=&\frac{1}{4\pi}\int_{0}^{\cos^{-1}\beta}P\cdot2\pi\sin\vartheta\,d\vartheta\nonumber\\
&=&\frac{\varepsilon}{1-\beta^{2}}\int_{0}^{1-\beta}x^{2}\;dx=\frac{\varepsilon}{3}\frac{(1-\beta)^{2}}{1+\beta}. \label{eq:P-tot}
\end{eqnarray}

This result  differs from Terrell~\cite{terrell} in the limits of integration, and from~\cite{rindler,schlegel}. The total radiation pressure, \eqref{eq:P-tot},  tends to its classical value of $\varepsilon/3$ in the limit as $\beta\rightarrow0$, and vanishes in the limit as $\beta\rightarrow1$, which is completely comprehensible since light waves cannot exert a pressure on an object which is traveling at the same speed. Whereas Terrell~\cite{terrell} finds the same classical limit for the radiation pressure, he concludes that \lq\lq it becomes infinite for $\beta=1$\rq\rq, which is \eqref{eq:P-tot} under $\beta\rightarrow-\beta$, i.e., the mirror is approaching the radiation source. Curiously, the arithmetic average of forward and backward  pressures, 
\[\half\left[P_{\mbox{\tiny tot}}(\beta)+P_{\mbox{\tiny tot}}(-\beta)\right]=(\varepsilon/3)\frac{(1+3\beta^2)}{(1-\beta^2)},\] where $\varepsilon/3$, is the radiation pressure of incoherent thermal radiation at rest, is precisely what von Laue~\cite{laue} finds for the $xx$ component of the stress tensor  in an inertial frame moving in the $x$-direction.

The process of reflection changes the $e$-measure of the relative speed, $\bar{\beta}$, into $\bar{\gamma}$, where we now introduce bars over the symbols to distinguish their $e$-lengths from their $h$-lengths (unbarred). That is, 
\begin{equation}
\bar{\gamma}=\tanh\gamma\label{eq:iso}
\end{equation}
is now the corresponding segment of the Lobachevsky straight line in velocity space. When \eqref{eq:iso} approaches   $\cos\vartheta$, the angle of parallelism is reached and the radiation pressure will vanish [cf. eqn \eqref{eq:tanh}]. The relations between first- and second-order relativistic effects are~\cite{varicak}:
\begin{eqnarray}
\cosh 2(u/c) & = & \cosh^{2}(u/c)+\sinh^{2}(u/c)\nonumber\\
& = & \cosh\gamma=\frac{1+\bar{\beta}^{2}}{1-\bar{\beta}^{2}}=\Gamma, \label{eq:gamma-bis}\\
\sinh 2(u/c) & = & 2\sinh(u/c)\cosh(u/c)\nonumber\\
&=& \sinh\gamma=\frac{2\bar{\beta}}{1-\bar{\beta}^{2}}. \label{eq:mom}
\end{eqnarray}

With $\bar{\beta}=\tanh(\gamma/2)$  the relative speed, and  $\cos\vartheta=\tanh\delta$, the conservation of energy demands:
\begin{equation}
P\bar{\beta} =  \varepsilon(\cos\vartheta-\bar{\beta})
 - \varepsilon^{\prime}(\cos\vartheta^{\prime}+\bar{\beta}),\label{eq:W}
 \end{equation}
 which is given explicitly by:
 \begin{eqnarray}
\lefteqn{2\varepsilon\frac{\sinh^{2}(\delta-\gamma/2)}{\cosh^{2}\delta}\tanh(\gamma/2) } \label{eq:W-bis}\\
& = &\varepsilon\left\{\tanh\delta -\tanh(\gamma/2)\right.\nonumber\\
&-&\left.\frac{\cosh^{2}(\delta-\gamma)}{\cosh^{2}\delta} \left[\tanh(\delta-\gamma/2)+\tanh(\gamma/2)\right]\right\}.\nonumber
\end{eqnarray}
Equation \eqref{eq:W} expresses the fact that the difference in energy is equal to the work, $P\bar{\beta}$. In the limit as $\delta\rightarrow\infty$, we get the line of sight relation [cf. eqn \eqref{eq:P}]:
\[
P\bar{\beta} = 2\varepsilon\bar{\beta} e^{-\gamma} =  2\varepsilon\bar{\beta}\left(\frac{1-\bar{\beta}}{1+\bar{\beta}}\right).
\]

When radiation impinges on a forward moving mirror the wavelength of incident radiation is shortened by the amount proportional to $(1-\bar{\beta})$, while the reflected radiation is elongated by an amount proportional to $(1+\bar{\beta})$.  In fact, \eqref{eq:W} is the negative of what Pauli~\cite{pauli}  considers as Einstein's expression for the radiation pressure.  $P\bar{\beta}$ is the work that is required to move the mirror backwards.

Poynting~\footnote{Poynting~\cite[p. 25]{poynting} realizes that \lq\lq the total energy is in inverse proportion to the square of the wave-length, when the height and depth remain the same.\rq\rq\ He also claims that \lq\lq the energy of the motion is proportional to the square of the velocity.\rq\rq\ The two taken together place the velocity in inverse proportion to the wavelength, which can be taken as de Broglie's relation, stated some fourteen years prior to de Broglie.} very vividly describes pressure absorption as the ceasing of wave motion at a black surface, where the waves deliver up all their momentum. Since the waves~\cite[pp. 31-32]{poynting} 
\begin{quote} press against [the black surface] as much as they pressed against [the source] in being emitted$\ldots$ the pressure against [the black surface] is therefore equal to the energy density per cubic centimetre in the beam.
\end{quote}  If the source is moving forward at constant relative speed $\bar{\beta}$ the  work, $P\bar{\beta}$, is determined by the  one-way Doppler shift; that is,  the difference between the incident energy density per unit area per unit time, $\varepsilon(\cos\vartheta-\bar{\beta})$, and the energy absorbed by the black surface $\varepsilon^{\prime}\cos\vartheta^{\prime}$:
\begin{eqnarray*}
P\bar{\beta} &=& \varepsilon(\cos\vartheta-\bar{\beta})-\varepsilon^{\prime}\cos\vartheta^{\prime}\\
&=& \varepsilon\left\{\cos\vartheta-\bar{\beta}-\frac{(1-\bar{\beta}\cos\vartheta)^{2}}{1-\bar{\beta}^{2}}\left(\frac{\cos\vartheta-\bar{\beta}}{1-\bar{\beta}\cos\vartheta}\right)\right\}\\
&=&\varepsilon\left\{\tanh\delta-\tanh(\gamma/2)\right.\\
& & \left.-\frac{\cosh^{2}(\delta-\gamma/2)}{\cosh^{2}\delta}\cdot\tanh(\delta-\gamma/2)\right\}.
\end{eqnarray*}
This gives a radiation pressure:
\begin{equation}
P=\varepsilon\frac{\sinh^{2}(\delta-\gamma/2)}{\cosh^{2}\delta}=\varepsilon\frac{(\cos\vartheta-\bar{\beta})^{2}}{1-\bar{\beta}^{2}}, \label{eq:P/2}
\end{equation}
which is exactly \emph{half\/} of \eqref{eq:terrell}. It has the same form as \eqref{eq:lav}, since the latter can be expressed as:
\[P=2\varepsilon\frac{\sinh^{2}(\delta-\gamma)}{\cosh^{2}\delta}.\]The fact that the wavelength at which the radiation is absorbed is greater than that at which it is emitted by the source, i.e. $\lambda^{\prime}/\lambda=(1-\bar{\beta}\cos\vartheta)^{-1}$, means that less energy is absorbed than was emitted. This is true also for the two-way shift. The factor of $2$ has led to the confusion of whether to consider the radiation pressure reflected by a moving mirror as being a one- or two-way Doppler shift, or equivalently, as belonging to the Klein or Poincar\'e model of the hyperbolic plane.

\section{angle of parallelism and the vanishing of the radiation pressure}

Consider a circle  of unit radius in velocity space $\bar{\beta}<1$. It is also a hyperbolic circle with center $O$ and some hyperbolic radius $\beta$ whose value is \eqref{eq:u}.
This  defines the \lq distance\rq\ $u$ in terms of the logarithm of the cross-ratio. Now consider the right triangle that has an angle $\vartheta$ at the origin, as shown in  Fig. 5. 
\begin{figure*}[floatfix]
	\centering
		\includegraphics[width=0.50\textwidth]{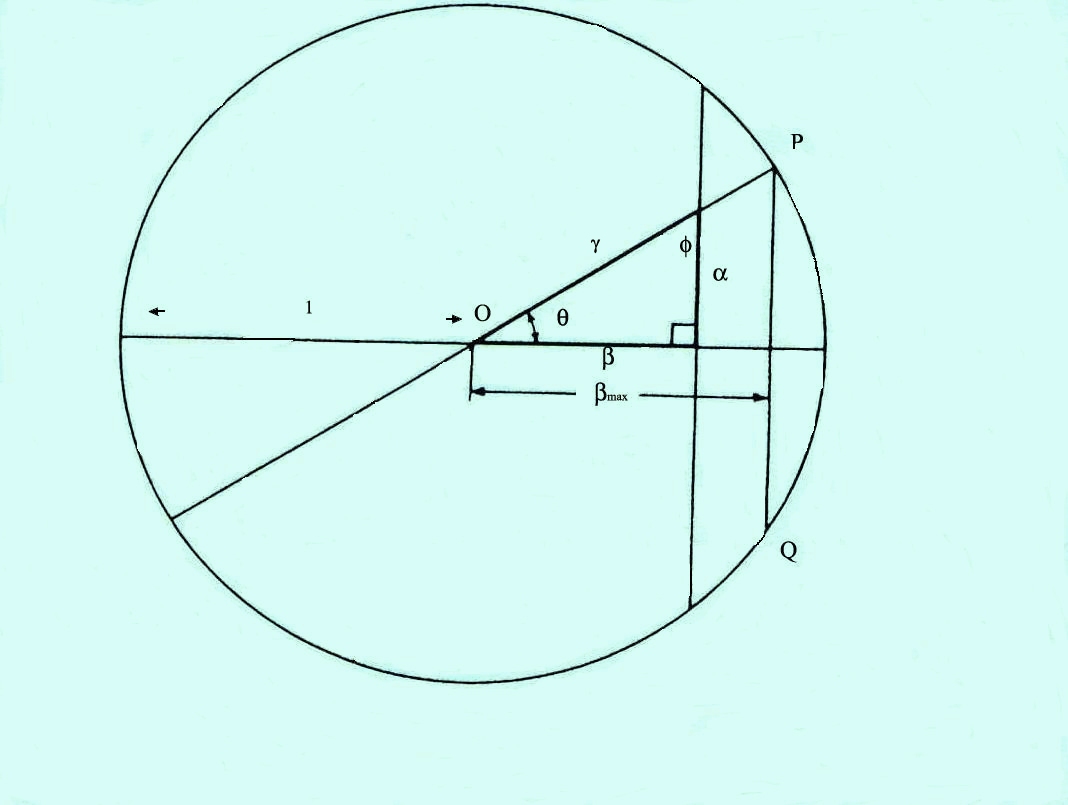}
	\caption{}
	\label{fig:P-5}
\end{figure*}

Since the angle $\vartheta$ is located at the origin, the $h$-measure of $\vartheta$ will be the same as its $e$-measure. Recalling that hyperbolic tangents correspond to straight lines in Lobachevsky space, the cosine of the angle will be the ratio of the adjacent to the hypotenuse,  $\cos\vartheta=\cos\bar{\vartheta}=\tanh\beta/\tanh\gamma$, where $\bar{\vartheta}$ is the $e$-measure of the angle, and we have set the absolute constant equal to one. We must now put a bar over $\beta$ to distinguish it from its $h$-measure $\beta$. Now, the $e$-length of the opposite side, $\bar{\alpha}$, can be calculated from the cross-ratio, and what is found is~\cite[p. 177]{busemann}: 
\begin{equation}
\bar{\alpha}=\tanh\alpha\;\mbox{sech}\beta. \label{eq:lorentz}
\end{equation} 
This is the origin of the Lorentz contraction in the direction normal to the motion~\cite{lavenda}. That is, $\bar{\alpha}$  is shortened by the amount $\mbox{sech}\beta=\surd(1-\bar{\beta}^2)$ from its $h$-measure $\tanh\alpha$ had it been located at the origin. This also responsible for the angle defect.

In the projective model, the $h$-measures of all other angles will be different than their $e$-counterparts. For the angle $\phi$ we have:
\[\cos\bar{\phi}=\frac{\bar{\alpha}}{\bar{\gamma}}=\frac{\tanh\alpha\;\mbox{sech}\beta}{\tanh\gamma}=\cos\phi\surd(1-\bar{\beta}^2).\]
Since the last term is less than unity $\cos\phi>\cos\bar{\phi}$. But since cosine is a decreasing function over the open interval $(0,\pi)$, it follows that $\phi<\bar{\phi}$, so that if the $e$-sum of angles of a triangle is $\pi$, its $h$-sum will be less than $\pi$. Thus, the angular defect is what is responsible for the Lorentz contraction in the direction normal to the motion~\cite{lavenda}. 

Now, the largest value of $\alpha$ occurs when it reaches the chord $PQ$. Its $h$-measure becomes infinite, and the angle $\phi$ tends to zero for $\beta\neq0$. However, the $e$-measure of $\alpha$ is:
\begin{equation}
\bar{\alpha}=\sin\vartheta(\baru_{\max})=\surd(1-\bar{\beta}^2_{\max})=\mbox{sech}\beta. \label{eq:alpha}
\end{equation} 
Only at the angle of parallelism can  rotation be linked to  $h$-contraction, and this is precisely what happens at $\vartheta=\cos^{-1}\bar{\beta}$. 

Since the $e$-measure of the hypotenuse, $\bar{\gamma}=1$, \eqref{eq:u} gives:
\begin{eqnarray*}
u_{\max} &= &
 \frac{c}{2}\ln\left(\frac{1+\cos\vartheta}{1-\cos\vartheta}\right)\\
 & = & \frac{c}{2}\ln\left(\frac{1+\cos\vartheta} {\sin\vartheta}\right)^2=c\ln\cot(\vartheta/2),
\end{eqnarray*}
where $\vartheta$ is the angle of parallelism, which is a  function of (half) the length $\baru_{\max}$, and $\bar{\beta}_{\max}=\cos\vartheta$ in Fig. 3. 

We recall that expression \eqref{eq:alpha}
is what Terrell~\cite{terrell-bis} finds for the rotation of an object that an observer will see in the same frame as the moving object when the stationary observer's view is in the direction normal to the motion. And since this is a limiting form of aberration it does not depend upon the distance between the observer and the object that is being observed. 

For a one-way shift, the radiation pressure \eqref{eq:terrell} vanishes at the critical angle $\vartheta=\cos^{-1}\bar{\beta}$, whereas for a two-way shift \eqref{eq:lav} vanishes at its critical angle, $\vartheta=\cos^{-1}\bar{\gamma}$. At these  critical angles, the waves have ceased to press against the mirror, and, consequently, the radiation pressure vanishes.  This is what the Klein disc predicts.

Now, let us see what the Poincar\'e disc has to say about two-way Doppler shifts. Since the model is conformal there is no need to distinguish between $e$- and $h$-measures of the angles. Consider  the  $h$-arc length, $\gamma$,  from $A$ to $B$ in the Poincar\'e half-plane, in Fig. 6, of a semi-circle of radius $1$.  Its length is determined by the logarithm of the cross-ratio $(A^{\prime}B^{\prime},PQ)$, where the primes denote the projections of  $A$  and $B$ onto the $x$-axis. 
\begin{figure*}[floatfix]
	\centering
		\includegraphics[width=0.50\textwidth]{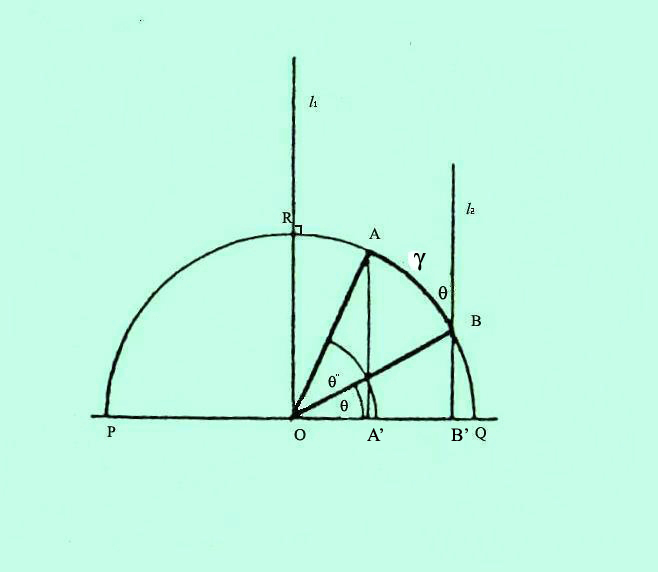}
	\caption{}
	\label{fig:P-6}
\end{figure*}
Using the Klein definition of $h$-distance, \eqref{eq:d-klein},  we have:
\begin{equation}
\gamma=\half\ln\left(A^{\prime}B^{\prime},PQ\right)=\half\ln\left(\frac{1+\cos\vartheta}{1-\cos\vartheta}\cdot\frac{1-\cos\vartheta^{\prime\prime}}{1+\cos\vartheta^{\prime\prime}}\right) ,\label{eq:s}
\end{equation}
where $\vartheta=\angle BOQ$ and $\vartheta^{\prime\prime}=\angle AOQ$.  Hence, the $e$-length of $\gamma$ is:
\begin{equation} \tanh \gamma=\frac{\cos\vartheta-\cos\vartheta^{\prime\prime}}{1-\cos\vartheta\cos\vartheta^{\prime\prime}},
\label{eq:Tanh}
\end{equation}
and when $\gamma$ becomes the $h$-length $RB$, $\vartheta^{\prime\prime}=\pi/2$, \eqref{eq:Tanh} reduces to [cf. eqn \eqref{eq:tanh}]:
\begin{equation}
\bar{\gamma}=\tanh\gamma=\cos\vartheta(\bar{\gamma}). \label{eq:iso-bis}
\end{equation}
Now, since
\begin{eqnarray}
\gamma & = & \tanh^{-1}\bar{\gamma}  = \half\ln\left(\frac{1+\bar{\gamma}}{1-\bar{\gamma}}\right)\nonumber\\
& = & \ln\left(\frac{1+\bar{\beta}}{1-\bar{\beta}}\right)=2\beta \label{eq:henri}
\end{eqnarray}
we, in effect, are dealing with Poincar\'e's definition of $h$-distance [cf. eqn \eqref{eq:u}].

Thus, $\vartheta$ becomes the angle of parallelism, which is a function solely of the arc length, $\bar{\gamma}$.  This is so because $BR$ is perpendicular to the line $\ell_{1}$ whose bounding parallel through $B$ is  $\ell_{2}$. Hence, the angle between $BR$ and $\ell_{2}$ is also equal to $\vartheta$.

\section{transverse Doppler shifts as experimental evidence for the angle of parallelism}

The  one-way Doppler shift, \eqref{eq:doppler}, predicts a small \lq blue-shift\rq\ when $\vartheta=\pi/2$,
\begin{equation} 
\nu^{\prime}=\nu/\surd(1-\bar{\beta}^{2}) \label{eq:blue1}. 
\end{equation}Ives and Stilwell~\cite{ives} were the first to test time dilatation by measuring the difference in the Doppler shift of spectral lines emitted in the forward and backward directions by a uniformly moving beam of hydrogen atoms. 

It might be more advantageous to consider the  two-way Doppler shift, where \eqref{eq:lambda-bis} gives the frequency shift:
\begin{equation}
\nu^{\prime\prime}=\nu\frac{(1+\bar{\gamma}\cos\vartheta)}{\surd(1-\bar{\gamma}^{2})}. \label{eq:doppler-bis}
\end{equation}
The two-way aberration formula,
\[
\sin\vartheta^{\prime\prime}=\frac{\surd(1-\bar{\gamma}^{2})}{1+\bar{\gamma}\cos\vartheta}\sin\vartheta, \]
together with \eqref{eq:doppler-bis}  lead immediately to the law of sines, \eqref{eq:mirror}, for a moving mirror, which, as we have point out, also happens to be the formula for aberration.

The two-way Doppler shift, \eqref{eq:doppler-bis}, like its one-way counterpart, \eqref{eq:doppler}, predicts a \lq red-shift\rq\ as either the transmitter or receiver recede from the other. However, for $\vartheta=\pi/2$, a blue-shift would remain. The shifted frequency would be
\begin{equation}
\nu^{\prime\prime}=\left(\frac{1+\bar{\beta}^{2}}{1-\bar{\beta}^{2}}\right)\nu=\nu/\sin\vartheta^{\prime\prime}. \label{eq:blue}
\end{equation}
In this limit, \eqref{eq:tan-bis} reduces to the angle of parallelism:
\begin{equation}
\tan(\vartheta^{\prime\prime}/2)=\left(\frac{1-\bar{\beta}}{1+\bar{\beta}}\right)=e^{-\gamma}, \label{eq:aop-bis}
\end{equation}
which follows from \eqref{eq:henri}. The angle $\vartheta^{\prime\prime}$ is, indeed, acute, and $\bar{\gamma}=\tanh\gamma=\cos\vartheta^{\prime\prime}$. Therefore, a second-order shift predicted by \eqref{eq:blue} would be a direct confirmation that relativity operates in hyperbolic velocity space. At the present time, the experimental evidence is not conclusive. Light pulses reflected from a rotating mirror have not shown relativistic frequency shifts~\cite{davies}, nor have those from dual disks rotating at equal speeds in opposite directions operating in the microwave region~\cite{thim}. However, a positive result has been reported by measuring the M\"ossbauer effect with source and absorber mounted on a rotating disk~\cite{champ}.

The null results can possibly be explained by a confusion between one-way and two-way Doppler shifts.  In~\cite{thim}, the one-way, \eqref{eq:blue1}, and two-way, \eqref{eq:blue}, shifts were placed on equal footing because both predict a frequency shift proportional to $\bar{\beta}^{2}$. Hence, it is not clear to experimenters what they should be looking for is a two-way, second-order Doppler shift, and not a first-order one.

\end{document}